\documentclass{article}

\newcommand{\define}{\emph}      % First/defining use of a phrase.
\newcommand{\foreign}{\textit} % Foreign phrase.
\newcommand{\latin}{\foreign}  % Latin phrase.

\newcommand{\ie}  {\latin{i.e.}}   % in other words
\newcommand{\eg}  {\latin{e.g.}}   % for example
\newcommand{\etc} {\latin{etc.}}   % et cetera (and so on)

\usepackage{amssymb}
\usepackage{amsthm}

\theoremstyle{definition}
\newtheorem{defn}{Definition}[section]

\title{The Indefinite Logarithm, Logarithmic Units, and the Nature of Entropy}
\author{Michael P. Frank\\
FAMU-FSU College of Engineering\\
Dept. of Electrical \& Computer Engineering\\
2525 Pottsdamer St., Rm. 341 \\
Tallahassee, FL 32310 \\
{\tt mpf@eng.fsu.edu}}

\date{\today}

\begin{document}
\maketitle
\begin{abstract}
We define the \define{indefinite logarithm} $[\log x]$ of a real
number $x>0$ to be a mathematical object representing the abstract
concept of the logarithm of $x$ with an {\emph{indeterminate}} base
(\ie, not specifically $e$, 10, 2, or {\em any} fixed number).  The
resulting indefinite logarithmic quantities naturally play a
mathematical role that is closely analogous to that of dimensional
physical quantities (such as length) in that, although these
quantities have no definite interpretation as ordinary numbers,
nevertheless the {\emph{ratio}} of two of these entities is
naturally well-defined as a specific, ordinary number, just like the
ratio of two lengths.  As a result, indefinite logarithm objects can
serve as the basis for {\define{logarithmic spaces}}, which are
natural systems of logarithmic units suitable for measuring any
quantity defined on a logarithmic scale.  We illustrate how
logarithmic units provide a convenient language for explaining the
complete conceptual unification of the disparate systems of units
that are presently used for a variety of quantities that are
conventionally considered distinct, such as, in particular, physical
entropy and information-theoretic entropy.
\end{abstract}

\section{Introduction}
The goal of this paper is to help clear up what is perceived to be a
widespread confusion that can found in many popular sources
(websites, popular books, \etc) regarding the proper mathematical
status of a variety of physical quantities that are conventionally
defined on logarithmic scales.

As an example of a logarithmic quantity about which much confusion
still lingers, we focus on the quantity of thermodynamic entropy,
and its close relationship to (and really, identity with) the
concepts of entropy and information as defined within the context of
information theory. Although many physicists and information
theorists have understood quite well the mathematical reasons for
the underlying unity between the entropy concepts in these two
domains, many others still do not, and continue to believe that the
physical and information-theoretic concepts of entropy are somehow
fundamentally different from each other.

Some (although not all) of the confusion that we have seen expressed
in this regard can be traced back to the historical accident that
thermodynamic entropy is most often measured in natural logarithm
units, while information-theoretic entropy is more frequently
measured in units of the logarithm base 2 (\ie, in bits), or some
multiple thereof.  But of course, the choice of the logarithm base
in the definition of entropy is completely inessential, and amounts
merely to a choice of one's unit of measurement, which went without
saying in Boltzmann's era, and which Shannon himself pointed out in
his seminal work {\cite{Shannon-48}} on information theory.

Even further, the supposed distinction between the ``physical''
nature of thermodynamic entropy (as measured in, say, Joules per
Kelvin) and the allegedly more ``mathematical'' nature of
information entropy (measured in bits) can also be seen as a totally
artificial distinction, one resulting from nothing other than the
fact that early thermodynamicists were not yet aware that physical
entropy really {\emph{is}} nothing other than a measurable
manifestation of what is at root merely a purely mathematical,
statistical quantity.

In fact, as we will review, any given unit of physical entropy can
be {\emph{exactly}} identified with a corresponding (purely
abstract) {\emph{mathematical}} unit, while still remaining
consistent with all observed empirical data.  The fundamental
scientific principle of adopting the most parsimonious theory that
explains the data (a.k.a. Ockham's razor) then {\emph{demands}} that
as good scientists we {\emph{must}} indeed adopt this identification
between the physical and mathematical domains, and take it seriously
as holding the status of our best available model of reality, at
least until empirical evidence to the contrary is found.

Although these issues are already quite well understood in certain
circles, we nevertheless felt that, as a public service, it would be
worthwhile to compose a short paper that elaborates on the
mathematical foundations of these issues in some detail.  Two
fundamental mathematical concepts which I have to be found rather
useful in explaining these kinds of issues are concepts that I refer
to as the ``indefinite logarithm'' and ``logarithmic units.'' The
definition and discussion of these concepts will form the main
mathematical core of this paper.

Although this material seems to be already essentially common (or
intuitively obvious) knowledge among many of the leading researchers
who deal every day with the physics of information, I have found in
my experience that misunderstandings and confusion regarding these
issues nevertheless still abound in other communities.

The reader should please note that, since this material seems to
hold the status of being considered obvious or common knowledge in
certain circles, this paper is by no means intended to claim any
kind of intellectual priority on these issues.  Rather, it is being
written simply because the author is not presently aware of an
accessible reference on this subject that explicitly explains these
issues with a sufficient degree of pedagogical detail to satisfy
general audiences.

The author welcomes comments and feedback from readers that may help
point the author at seminal references or review articles in the
mathematical literature that may elucidate these same issues, though
quite possibly using different terminology.

\section{The Indefinite Logarithm}

We use the standard notation $\log_b a$ for the logarithm, base $b$,
of $a$.  Of course, this expression is well-defined for all real
$a,b>0$, and even for all complex $a,b\neq 0$ as a multi-valued
function.  But, what if {\em no} particular base $b$ is selected?

Of course, as a matter of notational convenience in mathematical
literature, $\log a$ is often defined to be simply a shorthand for
the frequently-used natural logarithm, $\ln a = \log_e a$, or, in
more everyday applied contexts, for the decimal-based $\log_{10} a$.
But the topic of this paper is {\em not} situations such as these in
which some definite base really exists but is merely left implicit
by the notation. Rather, here we would like to discuss the concept
of a `new' kind of logarithm function wherein {\emph{no specific
base is implied at all}}.  We dub this the {\define{indefinite
logarithm}}, and we will give it a formal definition in a moment.

Of course, without a specific base, the result of the logarithm
cannot be an ordinary number, since any specific numeric result
would imply some specific base that must have been used.  Instead,
we can decare the output of the indefinite logarithm to be a
different (\ie, non-numeric) type of mathematical object
representing the result of performing this more abstract operation.

The form of this new type of object can be rigorously defined using
standard mathematical concepts.  Of course, as with any type of
mathematical object, there is an infinite variety of ways in which
we could satisfactorily represent these new objects in terms of more
standard mathematical objects. Here is the representation that we
find most convenient for purposes of this paper:

\begin{defn}[Indefinite logarithm]
For any given real number $x>0$, the {\define{indefinite logarithm
$L$ of $x$}}, written $L=[\log x]$, is a special type of
mathematical object called a ``logarithmic quantity'' object, which
we define as follows:
\begin{equation}
L = [\log x]:\equiv \lambda (b>0).\log_b x = \{(b,y)|b>0,\ y=\log_b
x\}.
\end{equation}
\end{defn}

Here, $:\equiv$ denotes ``is defined as,'' and in the first
expression after the $:\equiv$ we are using Church's lambda-calculus
notation for functions (see, for example, {\cite{Barendregt-84}}),
which gives us a concise way of saying that the indefinite logarithm
$[\log x]$ for any given $x$ is defined to be the unary function
object $L:\mathbb{R}\rightarrow\mathbb{R}$ mapping real numbers
$b>0$ to the logarithm $y=\log_b x$ of the (given constant) $x$ to
the (variable, function argument) base $b$. Meanwhile, on the right,
we are merely writing out the standard ``graph representation'' of
this function object explicitly as the set of all ordered pairs
$(b,y)$ consisting of a base $b>0$, followed by the ordinary
(definite) logarithm $y=\log_b x$, of $x$ to the base $b$.

Clearly, the result of the indefinite logarithm, as defined above,
does not select any preferred base, yet it contains ``all of the
information'' about the logarithm of $x$ taken to all possible
bases.  So, in this sense, it is not any less descriptive than a
definite logarithm.

Although the above definition is restricted to positive real numbers
(since this all that we need for our subsequent discussions), it
could easily be extended to non-zero complex numbers if desired.

We can define addition of indefinite logarithms by adding their
corresponding $y$ values:

\begin{defn}[Sum of indefinite logarithms]
Given two indefinite logarithm objects $L_1$ and $L_2$, their sum
$L=L_1+L_2$ is defined by $L(b) :\equiv L_1(b)+L_2(b)$ for all
$b>0$.  Or, stated a bit more formally using Lambda calculus, $L
:\equiv \lambda b.L_1(b)+L_2(b)$.
\end{defn}

We can similarly define negation of indefinite logarithms by simply
negating their $y$ values:

\begin{defn}[Negation of indefinite logarithms]
Given an indefinite logarithm object $L$, its negative $L'=-L$ is
defined by $L'(b) :\equiv -L(b)$ for all $b>0$.  Or,
$L':\equiv\lambda b.{-L(b)}$.
\end{defn}

If we add any indefinite logarithm to its negation, or take the
indefinite logarithm of 1, we get a unique indefinite-logarithm
object called the {\define{null}} or 0 indefinite logarithm, which
returns 0 for all bases:

\begin{defn}[Null indefinite logarithm]
The indefinite logarithm of 1, \ie\ $L_0=[\log 1]$, is (by previous
definitions) the function $\lambda b.0$ over reals $b>0$.  This
$L_0$ will be called the {\define{null indefinite logarithm}} and
will sometimes be written $[0]$.
\end{defn}

Of course, [0] is the identity element for the addition operation on
natural logarithms; that is, for any $L$, we have $L+[0]=L$.

Note that there is {\em no} corresponding concept of a {\em unit}
indefinite logarithm, \ie, a multiplicative identity.  That is,
indefinite logarithms are inherently {\em scale-free} objects; that
is, they are non-scalar quantities.  Further, the space of
indefinite logarithms does not even need to be considered to be
closed under multiplication.  A meaningful multiplication operation
can be defined if a product of two indefinite logarithms is
considered to be a distinct type (similarly to how the product of
two lengths is an area), but we will not develop that here.

Consistently with all of the above definitions, and with the
ordinary definition of multiplication as repeated addition,
indefinite logarithms can also be multiplied by arbitrary (positive,
negative, or zero) real numbers:

\begin{defn}[Indefinite logarithms multiplied by scalars]
Given an indefinite logarithm quantity $L$ and real number $r$,
define the product $L'=rL=Lr$ of $L$ times $r$ by $L'(b) :\equiv
r\cdot L(b)$. That is, $L'=\lambda b.rL(b)$.
\end{defn}

Finally, solving the preceding expression for $r$ allows us to
recognize and define the result of the ratio of two indefinite
logarithms as being an ordinary number:

\begin{defn}[Ratio of indefinite logarithms]
Given two indefinite logarithms $L_1$ and $L_2$, their ratio
$r=L_1/L_2$ is defined as the real number $r:\equiv L_1(b)/L_2(b)$,
where $b$ is any positive real number.  (The value of $r$ does not
depend on $b$.)
\end{defn}

As an immediate consequence of the above definitions, the ratio of
the indefinite logarithms of two numbers $a$ and $c$ is simply
$[\log a]/[\log c]=\log_c a$, which (note) is the same as the ratio
$\log_b a/\log_b c$ of the definite logarithms of $a$ and $c$ to any
common base $b$.  Thus, to emphasize, {\emph{the ratio of two
logarithms is independent of what base we are working in, and thus
remains well-defined even for indefinite logarithms}}.

The above fact is important for our discussions in subsequent
sections.

It is also worth noting that the indefinite logarithm shares all of
the mathematically important properties of the ordinary logarithm
(aside from not being a number), including the following useful
identities:

\begin{itemize}
\item $[\log xy] = [\log x] + [\log y]$
\item $[\log x/y] = [\log x] - [\log y]$
\item $[\log x^y] = y[\log x]$.
\end{itemize}

Finally, it is useful to also define an {\define{indefinite
exponential}} function, which maps a given indefinite logarithm
object back to the unique real number of which it is the indefinite
logarithm.

\begin{defn}[Indefinite exponential]
For any indefinite logarithm object $L = [\log x]$, let {\define{the
indefinite exponential of $L$}}, written $[\exp L]$, be given by
simply $[\exp L] = x$.
\end{defn}

Another way to define $[\exp L]$, which is helpful when we are not
explicitly given the $x$ such that $L=[\log x]$, is simply to say
that $[\exp L]=b^{L(b)}$, where $b>0$ is any positive real number;
all such $b$ give the same value for $[\exp L]$.

\section{Logarithmic Quantities}

In the above, we occasionally referred to the indefinite logarithm
objects as ``quantities'' in order to anticipate what we will now
discuss, which is that indefinite logarithmic quantities (which we
defined as pure mathematical entities) behave formally in a way that
is exactly analogous to how dimensional {\emph{physical}} quantities
(such as length, time, and mass) behave.  Indeed, logarithmic
quantities can be viewed as ``natural'' dimensional quantities that
exist independently of any particular models of physics.

Even further, later we will argue that logarithmic quantities can be
understood as being the underlying essence behind certain quantities
(in particular, thermodynamic entropy and physical information) that
are frequently perceived as being ``physical'' rather than
mathematical in nature.  We will also argue that the question of
whether these quantities are ``really'' physical or mathematical
ones is an ill-posed one, being predicated on an entirely false
dichotomy that has no real meaning.

First, what do we mean by a {\emph{quantity}}, in general?
(Regardless, for now, of whether it is supposed to be
``mathematical'' or ``physical.'') For our purposes, a quantity is
an object selected from a set having a structure similar to that of
the real number system, but without any built-in unit, that is, with
no pre-ordained object to be designated ``1.'' In abstract algebra
terms, a set of quantities forms an (abstract) vector space over the
reals, with a definite 0 quantity, an addition operation, a negation
operation, and the ability to multiply by reals, but without a
predefined unit quantity, and without necessarily any assigned
meaning for a product of quantities.

A bit more generally and formally, we can define:

\begin{defn}[Quantity spaces]
Given any field $F$ (in the standard abstract algebra
{\cite{Fraleigh-89}} sense of ``field,'' \ie, a commutative division
ring with unity), a {\define{quantity space Q over F}} is simply a
vector space over $F$, that is, an Abelian group of objects to be
called {\define{quantities}}, which is closed under the additional
operation of multiplication by the (scalar) elements of $F$.
\end{defn}

Although quantity spaces, being vector spaces, may in general be
many-dimensional, in this article we will primarily work with
examples of quantity spaces that are only one-dimensional.

We can now define the concept of a {\define{logarithmic (quantity)
space}}.

\begin{defn}[Logarithmic spaces]
A {\define{logarithmic space}} (or \define{logarithmic quantity
space}) is a quantity space $Q$ over $\mathbb{R}$ in which each
quantity $q\in Q$ is identified with an indefinite logarithm object
$L$ (as defined in the preceding section), or with a series of
indefinite logarithm objects, in the case of multidimensional
logarithmic quantity spaces.  The vector addition, negation, and
scalar multiplication operations are identified with the
corresponding operations on the indefinite logarithms.  The null (0)
vector comes from the null indefinite logarithm.
\end{defn}

Now, a {\define{logarithmic quantity}} $q$ is simply a member of
some logarithmic quantity space $Q$.  By a {\define{scalar}}
logarithmic quantity or {\define{logarithmic scalar}}, we mean a
member of a one-dimensional logarithmic space.  Members of
$n$-dimensional logarithmic quantity spaces will be called
{\define{$n$-dimensional} logarithmic quantities}.

\section{Logarithmic Units}

Quantities in general (and logarithmic quantities in particular)
have the property that there is no natural, built-in {\em unit}
quantity that is automatically provided by the quantity space
itself. However, in any given quantity space $Q$, we can always {\em
choose} some arbitrary $u\in Q$ to be designated as a provisional
{\define{unit quantity}}, and then all quantities in $Q$ can be
described in terms of scalar multiples of that unit.  (For elements
of multidimensional quantity spaces, a series of multiples is
needed.) We may even have several different units $u_1, u_2, \ldots
\in Q$, and express quantities sometimes as multiples of $u_1$,
sometimes as multiples of $u_2$, \etc, and convert between
expressions utilizing different units by multiplying them by
appropriate conversion factors.

Of course, we are already familiar with these properties of
quantities from their use in ordinary physics, in which (for
example) spatial distances (and multi-dimensional displacement
vectors) are considered to be quantities, rather than just pure
numbers, and we can choose any number of units (meters, feet, \etc)
for expressing them.  Space itself (in the traditional continuum
description) does not have any natural ``unit length,'' only
arbitrary units that we chose by convention.\footnote{Emerging
theories of quantum gravity suggest that the Planck length
$\ell_P=\sqrt{\hbar G/c^3}$ (or some small multiple of it) may play
the role of a ``natural'' unit length, in some sense which is not
yet fully understood. Nevertheless, we are still free, if we wish,
to treat lengths as quantities that can be represented in arbitrary
units.} Other examples of commonly used physical quantities include
time, velocity, mass, and energy.  (Of course, there are many others
as well.)

Now, the primary observation of the previous section is that spaces
of indefinite logarithm objects (logarithmic spaces) provide exactly
the structure of quantity spaces; thus, we can represent all
one-dimensional indefinite logarithm objects as scalar multiples of
some arbitrarily chosen ``unit'' indefinite logarithm object.
Indefinite logarithm objects thus naturally have the same
mathematical status, in this sense, as do physical quantities.

\section{Logarithmic Scales}

Logarithmic quantities and units, in one guise or another, are of
course very widely used today, for quantifying a wide variety of
concepts in different fields of study.  Some examples include:
\begin{itemize}
\item Relative signal amplitudes or power levels, in physics and engineering.
\item Earthquake strength (Richter scale) in seismology.
\item Tonal intervals on a musical scale.
\item Entropy (in, as we will see, both the thermodynamic and information-theory
senses).
\item Information, in the information theory sense.
\end{itemize}

What is lacking presently, however, is the ubiquitous understanding
that all of these disparate types of quantities can be understood as
dealing with what is fundamentally the {\em same} underlying system
of logarithmic units, as we defined above. The various ``different''
logarithmic scales that are in use are really distinguished only by
different choices of terminology for discussing logarithmic
quantities, different sizes and names of the logarithmic units used
for expressing them, and the application of these units in
describing different domains of study.

To illustrate, let us now identify and name a variety of indefinite
logarithm objects that are popularly used as units in which
logarithmic quantities of interest are expressed in various fields.
This list is ordered from the smallest logarithmic unit to the
largest, emphasizing that logarithmic units (like numbers) are
comparable across domains.

\begin{itemize}
\item $\mathrm{cent} = [\log 2]/1,\!200$.  In music theory, the
\define{cent} is 1/100th of a minor second, or 1/1,200$^{\mathrm{th}}$ of an octave.
\item $\mathrm{m2} = [\log 2]/12$.  In music theory, the
\define{minor second} m2 is 100 cents or 1/12$^{\mathrm{th}}$ of an octave.
\item $\mathrm{M2} = [\log 2]/6$.  In music theory, the
\define{major second} M2 is 200 cents or 1/6$^{\mathrm{th}}$ of an octave.
\item $\mathrm{dB} = 0.1[\log 10]$.  The \define{decibel}.  This is
the smallest logarithmic unit in widespread use outside music
theory, usually for expressing the magnitude of the ratio between
signal strengths.
\item $\mathrm{b} = [\log 2]$.  In information theory, the binary digit or \define{bit}.  This is the
smallest non-null logarithmic unit with an integer argument.  In
music theory, the same logarithmic unit is called an
\define{octave} P8.
\item $\mathrm{n} = [\log e]$.  The natural-log unit or \define{nat}.  As we will explain in more detail later,
this mathematical unit can be exactly identified with the physical
unit $k_\mathrm{B}$ known as Boltzmann's constant.  When used to
express a ratio of current or voltage levels, the nat is called a
\define{Neper} or Np.
\item $\mathrm{Np} = 2[\log e]$.  The magnitude of the \define{Neper}
of a ratio of currents or voltages, when translated to a ratio of
power levels. (It is doubled because the power is the square of the
current or voltage.)
\item $\mathrm{o} = [\log 8] = 3\,\mathrm{b}$.  The octal digit, which could be abbreviated \define{oit} (in analogy with {\em bit}).  It is equal to three bits.  Used as an information unit in computer engineering.
\item $\mathrm{d} = [\log 10]$.  The decimal digit, abbreviable as \define{dit}.  In various contexts, this unit is also known as
\define{Bel},
\define{power of ten},
\define{order of magnitude}, \define{Richter-scale point}, or
\define{decade}.
\item $\mathrm{h} = [\log 16] = 4\,\mathrm{b}$.  In computer engineering, the hexadecimal
digit is a unit of information, which might be called a
{\define{hit}}, but in practice, it is called a \define{nibble} or
\define{nybble}.
\item $\mathrm{B} = [\log 256] = 8\,\mathrm{b} = 2\,\mathrm{h}$.
The usual definition of a \define{byte} in computer engineering;
sometimes called an \define{octet} in network engineering.
\item $\mathrm{kcal}/\mathrm{mol}/\mathrm{K} \approx 503.6\,k_\mathrm{B}/\mathrm{molecule}
\approx [\log (4.9\times 10^{218})]/\mathrm{molecule}$. In
chemistry, the
\define{kilocalorie per mole per degree Kelvin} is a common intensive unit of
thermodynamic entropy, equivalent to about $503.6\,k_\mathrm{B}$ (or
nats or Nepers) per molecule.
\item $\mathrm{kb} = 1,\!000\,\mathrm{b} = [\log 2^{1000}]$.  An information unit known as
a \define{kilobit} in telecommunications.
\item $\mathrm{kb} = 1,\!024\,\mathrm{b} = [\log 2^{2^{10}}]$.  An information unit called
a \define{kibibit}, also known as a \define{kilobit} in computer
engineering.
\item Multiplying the above definitions by 8 gives the standard definitions for the
\define{kilobyte} unit of information, in the telecommunication and
computer-engineering contexts respectively.
\item Similarly for higher powers of 1,000 (or 1,024), with prefixes mega- (M), giga- (G),
tera- (T), peta- (P), exa- (E), zetta- (Z), and yotta- (Y).
\item $\mathrm{J}/\mathrm{K}\approx 7.243\times 10^{22} k_\mathrm{B} \approx 11\,\mathrm{ZB} \approx [\log 10^{3.14558\times 10^{22}}]$.  In thermodynamics,
the \define{Joule per Kelvin} is a common extensive unit of bulk
thermodynamic entropy.  Converted into information units, it is
about 11 zettabytes, meaning $11\times 1,\!024^7\,\mathrm{B}$.
\item $\mathrm{kcal}/\mathrm{K} = 4186.8\,\mathrm{J}/\mathrm{K}  \approx 3.03\times 10^{26}
k_\mathrm{B} \approx 45.2\,\mathrm{YB}$.  In chemistry, the
{\define{kilocalorie per degree Kelvin}} is a common extensive unit
of bulk thermodynamic entropy.  In information units, it is about 45
yottabytes, meaning $45\times 1,\!024^8\,\mathrm{B}$.
\end{itemize}

Of course, one could systematically define and name still larger (or
smaller) logarithmic units by applying larger (or smaller)
order-of-magnitude prefixes to the above.

The point of this exercise is to emphasize that all of the supposed
disparate logarithmic scales that are in use in these various fields
are ultimately all just different views of the {\em same}
fundamental logarithmic scale.  The various quantities and units
discussed on all of these logarithmic scales are all exactly
comparable with each other (with the exception of intensive units
such as kcal/mol/K, which are only comparable if specific quantity
of material is chosen, \eg 1 molecule in the above).

Among the logarithmic quantities that are in widespread use, perhaps
the quantity whose status {\em as} a logarithmic quantity is least
widely appreciated in some circles is the quantity known as
{\define{thermodynamic entropy}}.  Reviewing why this ``physical''
quantity is indeed, at root, truly just a logarithmic quantity is
the subject of the next section.

\section{Logarithmic Units and Entropy}

The original definition of the quantity known as {\define{entropy}},
first introduced by Rudolph Clausius in the mid-1800s
{\cite{Clausius-1865}}, was (in differential form)
\begin{equation}
\mathrm{d}S = \mathrm{d}Q/T
\end{equation}
where $\mathrm{d}Q$ represents an infinitesimal increment of heat
energy added to or removed from a system, and $T$ is the temperature
of the system.  Although this is only a differential definition, we
can presume that a physical system has a property called its total
entropy $S$, changes of which correspond to the increments
$\mathrm{d}S$.  (However, the original definition did not specify
the base value of $S$ for any particular cases.)

Clausius observed that in any thermodynamic process, the total
entropy (as he defined it) never decreased, since heat always moved
spontaneously from higher-temperature systems to lower-temperature
ones, and never vice-versa.  He postulated that the principle of the
non-decrease of entropy could be introduced as a fundamental law of
physics (``second law of thermodynamics''), equivalent to the other
(pre-existing) versions of the second law (impossibility of
perpetual motion machines, \etc).

Now, {\em prima facie}, Clausius' entropy does not seem in any way
to be a logarithmic quantity.  But, with the subsequent development
of statistical mechanics by Maxwell {\cite{Maxwell-1871}}, Boltzmann
(see {\cite{Cercignani-98}}), and Gibbs {\cite{Gibbs-06}} in the
late 1800s, the thermodynamic entropy came to be understood as
really being a statistical quantity that is naturally defined on a
logarithmic scale. The ``Boltzman'' form of the definition of
entropy (which evolved gradually from the $H$ quantity originally
defined by Boltzmann in {\cite{Boltzmann-1872}}) was expressed as
\begin{equation}\label{eq:Boltz}
S = k_\mathrm{B} \ln W,
\end{equation}
where $W$ denoted the number of possible distinct microscopic ways
of arranging the system (consistently with its macroscopic
description), and $k_\mathrm{B}$ denoted a fundamental entropy unit
first used by Planck (according to {\cite{Cercignani-98}}) which
came to be called {\define{Boltzmann's constant}}, which had a value
that (in conventional units of heat over temperature) was found to
be equal to about $1.38\times 10^{-23}\,\mathrm{J}/\mathrm{K}$.

Now, the traditional stance as to the status of this equation, which
is maintained today by many of the more traditional-minded
thermodynamicists, is that the entropy $S$ is fundamentally a
``physical'' quantity, namely a ratio of heat to temperature, and
Boltzmann's equation (\ref{eq:Boltz}) predicts what the value of
this quantity will be as a multiple of the Boltzmann's constant
unit, where the multiplier is the pure number obtained from $\ln W$.

However, what is arguably the preferred (simpler and more modern)
perspective on Boltzmann's equation is that it is merely a way of
rendering (in traditional units) the more elegant and fundamental
relation
\begin{equation}\label{eq:modernS}
S = [\log W],
\end{equation}
where now entropy is taken as being at root just an indefinite
logarithmic quantity, in the abstract sense that we outlined in the
previous sections.

The modern form (\ref{eq:modernS}) can be seen as being exactly
equivalent to (\ref{eq:Boltz}) if we simply declare that
\begin{equation}
k_\mathrm{B} = [\log e],
\end{equation}
since $[\log e]\ln W = [\log W]$.  Note, in particular, that the
choice of using $[\log e]$ as the unit in the original equation
(\ref{eq:Boltz}) was a completely arbitrary one, and was merely a
consequence of the choice of using the base-$e$ (``natural'')
logarithm in the formula.  So, we could equally validly re-render
eq.~(\ref{eq:Boltz}) in any of the following ways:
\begin{eqnarray}
S &=& k_\mathrm{b} \log_2 W \\
S &=& k_\mathrm{o} \log_8 W \\
S &=& k_\mathrm{d} \log_{10} W,
\end{eqnarray}
where $k_\mathrm{b} = [\log 2] = k_\mathrm{B}\ln 2$, $k_\mathrm{o} =
[\log 8] = k_\mathrm{B}\ln 8$, and $k_\mathrm{d} = [\log
10]=k_\mathrm{B}\ln 10$ are respectively binary, octal, and decimal
entropy units, corresponding to the bases of the logarithms used. Of
course, any of the other logarithmic units listed in the previous
section (and a continuum of other units as well) could also have
been used in Boltzmann's relation, with a suitable choice of
logarithm base.

Observe now that the relations $k_\mathrm{B} = [\log e]$ and
$k_\mathrm{B} \approx 1.38\times 10^{-23}\,\mathrm{J}/\mathrm{K}$
imply that $1\,\mathrm{K} \approx 1.38\times
10^{-23}\,\mathrm{J}/[\log e]$, in other words, the Kelvin (or any
temperature unit) is fundamentally just an expression of an amount
of energy per logarithmic unit of some arbitrary size.  Here it is
expressed as energy per nat or Neper, where this unit quantifies the
increase in the indefinite logarithm of the number of states when
the state count is multiplied by $e$. Of course, we could equally
well express the Kelvin in terms of logarithmic units of other sizes
as well, for example, multiplying top and bottom by $(\ln 10)$ gives
$1\,\mathrm{K} \approx 3.18\times 10^{-23}\,\mathrm{J}/[\log 10]$,
where we see we have now expressed the Kelvin in units of Joules per
decade (Bel, order of magnitude, power of ten \etc) of increase in
the number of states.

Indeed, the modern thermodynamic definition of temperature is indeed
just
\begin{equation}
T = \mathrm{d}Q/\mathrm{d}S,
\end{equation}
where $\mathrm{d}Q$ is the amount of heat that must be added to a
system in order to increase its entropy $S=[\log W]$ by a small
amount $\mathrm{d}S$, or in other words to increase its number of
states by the multiplicative factor $[\exp \mathrm{d}S]$, where note
we are using our indefinite exponential notation from earlier.

\section{Logarithmic Units and Information}

Just as with entropy, the amount of information content or
information capacity $I$ of a system can be expressed very elegantly
and generically as an indefinite logarithmic quantity, that is, as
\begin{equation}
I = [\log W]
\end{equation}
where $W$ is again the number of ways of arranging the system, or a
subsystem of it whose state can be controlled, \eg, for purposes of
storing or communicating a message.

The only real distinction between entropy and information is a
distinction of epistemological status.

Entropy is usually taken to refer to that part of the physical
information that is {\emph{unknown}}, or in other words is not
included in the available overall description of a physical
situation; this information is not considered part of the so-called
``macrostate'' of the system.

The word ``information,'' on the other hand, is sometimes reserved
to implicitly connote information that is (or could be) explicitly
known, although this more restricted usage is becoming less common.
More and more, physicists who use the word ``information''
understand that physical information, in general, could have the
status of being either known information, or unknown information
(entropy). One word that has been proposed to indicate known
information as opposed to entropy, but which is not yet very
popular, is \define{extropy}, which was coined to serve as a
complement to the word ``entropy.''

Of course, due to the (historically dominant) use of binary codes in
our modern digital systems, information has been traditionally
measured in $[\log 2]$ units (bits), or multiples thereof (such as
bytes), rather than in $[\log e]$ units (nats) or $[\log 10]$ units
(decades).  However, as we have been emphasizing, this difference in
the conventional choice of units does not at all imply that
information is fundamentally a different kind of quantity from the
logarithmic quantities that are used in other contexts.

In fact, we argue that the only distinction between
information/entropy and other types of logarithmic quantities is
that information is a logarithmic quantity derived from an absolute
pure number that represents a ``number of alternatives'' in some
sense, while most other logarithmic quantities (\eg\ octaves,
decibels, Richter scale points) are derived from pure numbers
representing {\em ratios} between physical quantities (pitch, signal
power, Earthquake strength).  But fundamentally, although the
sources of the pure numbers $x$ in the two kinds of cases are
different, this does not matter; the values of $[\log x]$ are always
still fundamentally the very {\emph{same}} type of mathematical
object.

Thus, a bit of information {\emph{is}}, mathematically, the very
{\emph{same}} kind of object as an octave of pitch.  A Boltzmann's
constant unit of entropy is the same kind of object as a Naper of
current ratio. A decimal-digit-sized quantity of information is the
{\emph{same}} as a Richter-scale point of relative earthquake
strength.  Fundamentally, the only import of the different names for
these mathematical objects is to connote their use in describing
different types of situations.

Just as the number 2 is still a 2 whether we are talking about two
giraffes or two potatoes, likewise the indefinite-logarithm object
$[\log 2]$ is still a $[\log 2]$ unit whether we are talking about a
$[\log 2]$ amount of information (called a bit) or a $[\log 2]$ size
of musical interval (called an octave). And a $[\log e]$ unit is
still a $[\log e]$ unit, whether we are talking about a $[\log e]$
unit of thermodynamic entropy (called Boltzmann's constant) or a
$[\log e]$ unit of voltage ratios (called a Neper).  A $[\log 10]$
unit is still a $[\log 10]$ unit, whether we are talking about a
$[\log 10]$ unit of signal power ratio (called a Bel) or a $[\log
10]$ unit of information (called a decimal digit).

In other words, logarithmic units are logarithmic units, and
logarithmic quantities (expressed by a real number times a
logarithmic unit) are mathematically always the very same kind of
entity, no matter the domain. The only differences are in the size
of the standard units that are conventionally used in a given
context, the names that we call them, and how we apply them.

\section{Discussion}

Today, thanks to Boltzmann and his followers, we know that a certain
quantity that used to be thought of as ``physical,'' namely entropy,
is really just a ``mathematical'' quantity, namely an indefinite
logarithmic quantity derived from the number of states, or in other
words a kind of information.  Since indefinite logarithmic units are
scale-free, there is no natural unit or ``atom'' of entropy or
information that we must use, only units that we choose rather
arbitrarily, by convention or for mathematical or technological
convenience, such as the nat (Boltzmann's constant) or the bit.

In the future, it is possible that we might discover that other
quantities that are currently thought of as ``physical'' could at
root turn out to really be logarithmic quantities as well.  For
example, Tommaso Toffoli has speculated {\cite{Toffoli-98}} that,
just as entropy turned out to be equivalent to the logarithm of the
number of possible states that a system could (statically) be in,
perhaps {\emph{energy}} could be shown in some way to really be
equivalent a logarithm of the number of possible
{\emph{computations}} that a system could carry out dynamically at
the microscale.  As of this writing, this intriguing idea is still
rather far from being substantiated, but the history of how our
understanding of the quantity of entropy has evolved indeed makes
Toffoli's proposal seem like an idea worth exploring.

Besides entropy and (perhaps) energy, one wonders whether other
kinds of physical quantities such as distances and times might
potentially be shown to ultimately be logarithmic quantities as
well.  That this might be true is hinted at by the
Bekenstein-Hawking formula {\cite{Bekenstein-72,Hawking-74}} for the
entropy of a black hole, $S=A/4$, where $A$ is the hole's event
horizon area in Planck units, and $S$ is the entropy in nats.  Thus,
for example, we could assign the Planck unit of length to be the
square root of a nat, $\ell_{\mathrm{P}} = [\log e]^{1/2}$, and then
write $A=[\log W^4]$, where $W$ is the number of states of the black
hole, and this would be consistent with the Bekenstein relation as
well as the entropy relation $S=[\log W]$. However, in this line of
thought, it remains obscure why the area should be the indefinite
logarithm of the {\em fourth} power of the number of states, and why
the length unit should have dimensions of a square root of a
logarithmic unit. Still, this may be an interesting line to pursue
further.

\section{Conclusion}

In this paper, we have reviewed a well-defined mathematical concept
of an {\em indefinite} logarithm function in which no particular
logarithm base is selected, and have shown that the entities
returned by this function can be used as the basis for a system of
mathematical quantities that is exactly analogous in its behavior to
systems of dimensioned physical quantities.  In fact, this
mathematical system of indefinite logarithmic quantities {\em
exactly} corresponds the physical quantity known as entropy, when
the logarithms are applied to the number of distinguishable physical
states that are consistent with a given abstract description of the
system. This quantity (the indefinite logarithm of the number of
states) is also called ``information'' in a slightly broader
context.

In other words, physical (thermodynamic) entropy really {\emph{is}}
nothing but (unknown) information in the physical state, and its
quantity really {\emph{is}} nothing other than the indefinite
logarithm of the state count. Further, Boltzmann's constant
$k_\mathrm{B}$ is really nothing other than a representation (in
conventional physical units of energy over temperature) of the
specific (and arbitrarily chosen) abstract indefinite-logarithm unit
$[\log e]$, which is known as the nat or the Neper in other
contexts. And, thermodynamic temperature really is nothing but the
energy per logarithmic unit, for small increments in the indefinite
logarithm of the state count.

There are speculations that other quantities such as energy that we
presently think of as being fundamentally ``physical'' in nature (as
opposed to mathematical) might (similarly to entropy) someday be
revealed to be, at root, derived from logarithmic quantities of some
sort.

Of course, if physics can someday be {\em exactly} described by
mathematics, as most theoretical physicists believe (or at least
hope), then ultimately, the entire distinction between mathematical
and physical quantities becomes somewhat of an artificial and
illusory one, since we cannot then rule out the possibility that our
physical universe may really {\em be} nothing but a particular (very
elaborate) mathematical structure, one in which we (and our thought
processes) happen to be embedded.  What is a ``physical'' quantity
then ultimately becomes only a question of which mathematical
quantities happen to arise naturally within the context of the
particular mathematical structures that make up our physical
universe.

To conclude, although there are probably no substantive ideas in
this paper that have not been said many times before, somewhere in
the literature (though perhaps in different terms), and although
many of these ideas would likely be considered self-evident to
professional mathematicians, we nevertheless felt that many of these
ideas lack exposure at present within certain communities, and that
it would be worthwhile to present and explain them again, so as to
facilitate the more widespread understanding of these issues.  We
hope that this paper serves that purpose, at least.

{\bf Note:} The reference list below is still under construction.
The author would appreciate receiving from readers suggested
references to appropriate prior sources that discuss these or
similar ideas, so that he can cite the sources in future versions of
this paper, as well as in future papers on related topics.

\bibliography{indef-log}

\begin{thebibliography}{10}

\bibitem{Shannon-48}
Claude~E. Shannon.
\newblock A mathematical theory of communication.
\newblock {\em Bell System Tech. J.}, 27:379--423, 623--656, 1948.

\bibitem{Barendregt-84}
Henk~P. Barendregt.
\newblock {\em The Lambda Calculus: Its Syntax and Semantics}.
\newblock Elsevier, 1984.

\bibitem{Fraleigh-89}
John~B. Fraleigh.
\newblock {\em A First Course in Abstract Algebra}.
\newblock Addison-Wesley, fourth edition, 1989.

\bibitem{Clausius-1865}
Rudolph Clausius.
\newblock \"{U}ber verschiedene f\"{u}r die anwendung bequeme formen der
  hauptgleichungen der mechanischen w\"{a}rmetheorie.
\newblock {\em Poggendorff's Annalen}, 125:353.

\bibitem{Maxwell-1871}
James~Clerk Maxwell.
\newblock {\em Theory of heat}.
\newblock Longmans, Green, London, 1871.

\bibitem{Cercignani-98}
Carlo Cercignani.
\newblock {\em Ludwig Boltzmann: The Man Who Trusted Atoms}.
\newblock Oxford University Press, 1998.

\bibitem{Gibbs-06}
H.A. Bumstead and R.G.~Van Name, editors.
\newblock {\em The scientific papers of J.~Willard Gibbs}.
\newblock Longmans, Green, New York, 1906.

\bibitem{Boltzmann-1872}
Ludwig Boltzmann.
\newblock Weitere studien \"{u}ber das w\"{a}rmegleichgewicht unter
  gasmolek\"{u}len.
\newblock {\em Sitzungsberichte der Akademie der Wissenschaften, Wien, {II}},
  66:275--370, 1872.
\newblock English translation in S.G. Brush, \emph{Kinetic theory}, Vol. 2,
  \emph{Irreversible processes}, pp. 88--175, Pergamon Press, Oxford, 1966.

\bibitem{Toffoli-98}
Tommaso Toffoli.
\newblock Action, or the fungibility of computation.
\newblock In Anthony~J.G. Hey, editor, {\em Feynman and Computation}, pages
  348--392. Perseus, 1998.

\bibitem{Bekenstein-72}
Jacob~D. Bekenstein.
\newblock Black holes and the second law.
\newblock {\em Lett. Nuovo Cimento Soc. Ital. Fis.}, 4:737, 1972.

\bibitem{Hawking-74}
Stephen~W. Hawking.
\newblock Black hole explosions.
\newblock {\em Nature}, 248:30, 1974.

\end{thebibliography}
\bibliographystyle{unsrt}       % Use citation order.

\end{document}